\documentclass[twocolumn]{article}
\usepackage{amsmath,amsfonts}
\usepackage[caption=false,font=normalsize,labelfont=sf,textfont=sf]{subfig}
\usepackage{url}
\usepackage{verbatim}
\usepackage{graphicx}
\usepackage[
    style=ieee,
    sorting=none,
    backend=biber,
    isbn=false
]{biblatex}
\usepackage{hyperref}
\hypersetup{hidelinks}

\usepackage{siunitx}

\providecommand{\keywords}[1]
{
  \small	
  \textbf{\textit{Keywords---}} #1
}

\begin{document}

\title{A Novel Two-Step Approach for Reactive Power Demand Calculation Using Integrated Voltage Stability Analysis}

%\author{ \href{https://orcid.org/0000-0002-6416-8238}{\includegraphics[scale=0.06]{orcid.pdf}\hspace{1mm}Hassan~Abouelgheit}\footnote{Corresponding author} \\
%	Institute of Energy Process Engineering and Dynamics in Energy Systems\\
%	University of Stuttgart\\
%	70569 Stuttgart\\
%	\texttt{hassan.abouelgheit@ied.uni-stuttgart.de} \\
	%% examples of more authors
%	\And
%	\href{https://orcid.org/0000-0002-0208-410}{\includegraphics[scale=0.06]{orcid.pdf}\hspace{1mm}Hendrik~Lens} \\
%	Institute of Energy Process Engineering and Dynamics in Energy Systems\\
%	University of Stuttgart\\
%	70569 Stuttgart\\
%	\texttt{hendrik.lens@ied.uni-stuttgart.de} \\
%}

\author{\href{https://orcid.org/0000-0002-6416-8238}{\includegraphics[scale=0.06]{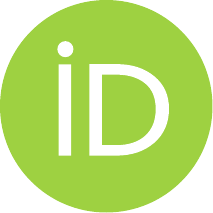}\hspace{1mm}Hassan~Abouelgheit}\footnote{Corresponding author,\\ \texttt{hassan.abouelgheit@ied.uni-stuttgart.de}} \footnote{
	Institute of Energy Process Engineering and Dynamics in Energy Systems, University of Stuttgart, 70569 Stuttgart} , 
	\href{https://orcid.org/0000-0002-0208-410}{\includegraphics[scale=0.06]{orcid.pdf}\hspace{1mm}Hendrik~Lens}\footnotemark[\value{footnote}]}

% Uncomment to remove the date
\date{}

% Uncomment to override  the `A preprint' in the header
%\renewcommand{\headeright}{Technical Report}
%\renewcommand{\undertitle}{Technical Report}
%\renewcommand{\shorttitle}{A Novel Two-Step Approach for Reactive Power Demand Calculation}
\maketitle

%\twocolumn[
%  \begin{@twocolumnfalse}
    \begin{abstract}
      The assessment of reactive power demand plays an instrumental role in power system planning. This paper presents a methodology for calculating reactive power demand based on a two-step approach. Unlike existing methodologies in the literature that focus primarily on optimization of reactive power compensation equipment placement and sizing through single-simulation approaches, this methodology directly calculates the actual reactive power demand through a comprehensive back-to-back simulation framework. While existing methods address either long-term or short-term voltage stability using either steady-state analysis or individual dynamic simulations, the proposed approach integrates both stability assessments sequentially through iterative Quasi-Dynamic Simulation, Q-V analysis and dynamic simulation. Furthermore, this methodology employs comprehensive time-series analysis over a full annual period (8760 hours) with multi-criteria violation assessment (number, severity and duration of voltage violations). In the final section of this paper, a case study was conducted to demonstrate the application of the proposed methodology. Simulations were performed to validate the effectiveness of the methodology, with the results showing that all buses with voltage issues were successfully addressed and finally the total reactive power demand across the network was calculated.

      \keywords{long-term voltage stability, short-term voltage stability, Q-V analysis, reactive power requirements, time series simulation, trajectory violation integral, dynamic simulation, power system dynamics, reactive power planning.}
%      \vspace{10pt}
    \end{abstract}
%  \end{@twocolumnfalse}
%]{
\section{Introduction}
\subsection{Background}

The global energy sector is experiencing a rapid transformation driven by sustainability goals and climate change mitigation efforts \cite{boricic_comprehensive_2021}. Countries worldwide are increasing the integration of Renewable Energy (RE) into their power grids to achieve net-zero carbon emissions, with ambitious targets such as Germany's objective of RE generation’s share of 80\% by 2050 \cite{iea_germany}. However, this transition presents significant challenges to transmission systems, mainly due to the massive integration of intermittent inverter-based resources (IBRs) such as wind and solar power generation plants \cite{fang_03_reactive_2014}. In contrast to traditional power plants, IBRs produce variable and not fully predictable output, which may lead to changing operating points causing variable power flow conditions that are challenging for the system, in particular with respect to reactive power balance and voltage stability. The simultaneous phasing out of conventional synchronous generators, which historically provided system inertia and reactive power support, further complicates these issues, fundamentally changing the dynamic behavior and stability of power systems \cite{dynamic_performance_committee_2024}.

According to \cite{hatziargyriou_definition_2021}, power system stability is classified into resonance, rotor angle, voltage, frequency, and converter-driven stability. The main focus of this research is the role of reactive power sources for voltage stability. The term \emph{voltage stability} is defined as \emph{the ability of a power system to maintain steady voltages close to nominal value at all buses in the system after being subjected to a disturbance}. This ability depends on the maximum power transfer and the availability of reactive power to compensate for the voltage drop resulting from power transmission or system faults. Voltage instability, primarily caused by reactive power deficiencies, can trigger voltage collapse, leading to cascading outages and widespread blackouts. The voltage stability of a system depends on the relationship between reactive power ($Q$) and voltage ($V$), where increased reactive power injections should produce voltage increases and vice-versa \cite{Kundur_Power_1994}. IBRs complicate this relationship through their diverse reactive power characteristics and intermittent nature, creating dynamic demands for reactive power compensation. Consequently, preparing electrical networks for future operating conditions with high IBR penetration is essential. Reactive Power Planning (RPP), which optimizes the type, location, and size of compensation equipments to ensure system security \cite{pranto_04_reactive_2021}, must incorporate voltage stability constraints. The uncertainties from intermittent renewables require new, robust methodologies to guarantee dynamic security in modern power grids.

The primary objective of this work is to develop a methodology to calculate the required reactive power demand to maintain voltage stability in a given power system. Only voltage phenomena related to transmission systems such as short circuit faults, changes in power flow due to changes in generation and load situations are considered. Fault Induced Delayed Voltage Recovery (FIDVR) events related to heating and cooling applications or motor overloading such as motor-stalling are not in the scope of this work.

\subsection{Related Work}

The increasing complexity of modern power systems driven by renewable energy integration and AC/DC hybrid networks has led to broad research on reactive power management and voltage stability. A systematic review of the existing literature reveals that prior works can be divided into three major categories based on how reactive power and voltage stability are addressed: \emph{(i)} studies that formulate reactive power planning as an optimization problem with voltage as a constraint; \emph{(ii)} studies that focus primarily on the development and application of voltage stability metrics and indices, with reactive power serving as an analytical rather than an optimized variable; and \emph{(iii)} studies that consider reactive power in a non-systematic, application-specific, or on device-level context, without a formal optimization framework or voltage stability analysis.

The most common approach in the literature, as in  \cite{barot_02_optimal_2007,fang_03_reactive_2014,diaz_07_reactive_2009,han_10_reactive_2022,pudjianto_11_der_2020,wildenhues_optimal_2015,xiaojiao_12_distribution_2022}, and \cite{guo_14_reactive_2021}, is to formulate reactive power planning or optimization as a constrained mathematical problem, where the goal is to determine the optimal size and location of reactive power sources, such as capacitor banks, SVCs, or generator reactive outputs, while keeping bus voltages within acceptable limits or close to predefined values in the sense of a chosen metric. The objectives defined across these works vary. Some minimize total operating or installation cost \cite{fang_03_reactive_2014,diaz_07_reactive_2009,wildenhues_optimal_2015}, others focus on reducing active power losses \cite{barot_02_optimal_2007, pudjianto_11_der_2020}, and some combine multiple goals \cite{han_10_reactive_2022, xiaojiao_12_distribution_2022}. Classical optimal power flow (OPF) methods are used, while metaheuristic approaches such as particle swarm optimization appear in \cite{han_10_reactive_2022}. More recent contributions such as \cite{yu_09_research_2022} and \cite{guo_14_reactive_2021} have turned to machine learning and reinforcement learning techniques to handle the complexity and computational efforts related to solving the problem. These studies operate under steady-state assumptions and partially address long-term voltage stability to ensure voltages remain within bounds as the system load gradually increases or decreases. Only one paper \cite{guo_14_reactive_2021} in this context specifically targets short-term voltage stability by responding to fault-induced voltage sags in real time.

Other research works such as \cite{jiang_12_quantification_2021,dondariya_15_voltage_2021,ziegler_16_voltage_nodate} and \cite{munyao_17_voltage_2021} do not focus on optimizing reactive power, but on understanding and assessing voltage stability itself. The papers in this group develop or apply voltage stability indices to assess how close a system is to the point of voltage collapse, using reactive power as a tool for analysis rather than a variable to be controlled or determined. For instance, \cite{jiang_12_quantification_2021} derives an energy-function-based stability index that accounts for transformer tap changer effects, while using continuation power flow analysis to trace PV curves and identify critical loading points in systems with increasing solar PV penetration. In \cite{ziegler_16_voltage_nodate}, a Thévenin-equivalent model is used to define a Power Transfer Stability Index (PTSI), and reactive power sensitivities are computed to guide operator decisions. Similarly, \cite{munyao_17_voltage_2021} combines Q-V curve analysis with eigenvalue decomposition to produce a multi-bus voltage stability index. All of these works operate in the static, long-term voltage stability domain. None of these publications involves dynamic or transient simulation.

The third group of references, \cite{song_01_comprehensive_2023,pranto_04_reactive_2021,jayasudha_05_sizing_2025,zhang_06_a_2009,du_08_discussion_2010,li_18_dynamic_2020,satyamsetti_19_active_2021} and \cite{kang_20_research_2025}, is more diverse and includes works where reactive power is addressed in a practical or engineering context, without systematic optimization or a formal stability analysis. Some of these works propose configuration guidelines for reactive power compensation across multiple voltage levels, or present novel compensation schemes tailored to specific network architectures such as ultra-high-voltage AC transmission lines. Others size compensation devices based on grid code compliance requirements or use post-fault voltage sensitivity rankings to guide device placement in a heuristic manner. Reference \cite{du_08_discussion_2010} provides a conceptual discussion of the challenges in dynamic reactive power planning without offering a concrete solution or case study. At the device level, \cite{kang_20_research_2025} demonstrates the STATCOM compensation principle through a laboratory circuit and focuses entirely on testing harmonic suppression performance of low-voltage compensation devices under standardized conditions. Where voltage stability is mentioned in this group, it tends to appear as a broad motivation rather than a formally analyzed property.

\subsection{Motivation}
While existing methods addressing reactive power compensation predominantly formulate the problem as an optimization task mainly with respect to investment costs or equipment placement, methods that consider voltage stability regard either long-term or short-term voltage stability, lacking an integrated framework that considers both stability aspects together. Moreover, these methodologies fail to provide a direct method to calculate the required reactive power for a specified stability margin.

This paper focuses on voltage stability and proposes a methodology to directly determine the necessary reactive power compensation across both stability aspects and is structured in three more sections. Section \ref{sec:methodology} provides a detailed overview of the novel concept and outlines the processes involved in each step as well as the assessment tools used in the methodology. In Section \ref{sec:study_case}, the concept is validated using a study case and the results are discussed. Finally, Section \ref{sec:conclusion} concludes this work and provides an outlook on potential future work related to the topic.

\section{Methodology}
\label{sec:methodology}
\begin{figure}
  \centering
  \includegraphics[width=\columnwidth]{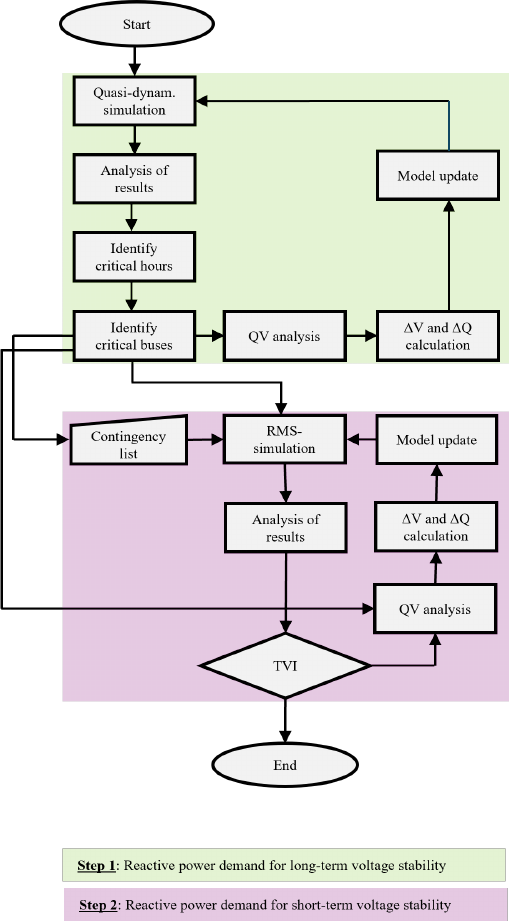}
  \caption{Proposed two-step approach for reactive power demand calculation}
  \label{fig_Method overview}
\end{figure}

Figure~\ref{fig_Method overview} shows a flow chart that represents the methodology proposed in this research work. The methodology calculates the reactive power demand of a network by a two-step approach that involves multiple simulations and analyses.

\subsection{Description}
\subsection*{Step 1: Long-term Voltage Stability Assessment}
The first step calculates the required reactive power demand for long-term voltage stability by performing a time-series (quasi-dynamic) simulation (QDS). In this work, the system was analyzed over a one-year period on an hourly basis (8760 hours). The simulation results are analyzed against a pre-defined voltage band  
\begin{equation}
\mathcal{V} = \left[ V_{\min}, V_{\max} \right]
\end{equation}
and violation criteria quantified through the number ($f_n$), severity ($\Delta V$), and duration ($t_{v}$) of voltage violation to this pre-defined voltage band. Following this analysis, critical time periods and buses where violations occur are identified. The application of these criteria eliminates buses without voltage violations and focuses only on buses experiencing voltage violations. Critical buses are then ranked according to these criteria, creating a prioritized list for remedial action.

Once all critical buses are identified and ranked, a Q-V analysis is performed, beginning with the highest-priority bus, to calculate the reactive power demand \emph{($\Delta${$Q_{i}$})} required to compensate for the voltage violation \textit{($\Delta${$V_{i}$})} at that specific bus. A reactive power compensation equipment is then installed at the identified bus location, followed by another QDS to assess the system's improved performance and possible new voltage violations. This iterative process continues through all critical buses from each iteration until all voltage violations are resolved. The total reactive power demand for long-term voltage stability is determined as 
\begin{equation}
\mathbf{\Delta Q}_{\mathrm{long}} =
\begin{bmatrix}
\Delta Q_{\mathrm{long},1}\\
\Delta Q_{\mathrm{long},2}\\
\vdots\\
\Delta Q_{\mathrm{long},N}
\end{bmatrix}
\end{equation}
where $i\in \{1, \dots, N\}$ is the bus number.

\subsection*{Step 2: Short-term Voltage Stability Assessment}
The results from Step 1 serve as input for Step 2, where only the critical timestamps (hours) at which minimum voltages in the system occur, both before and after compensation from Step 1, are considered for further analysis. In Step 2, a contingency list is defined and dynamic (RMS) simulations are performed for each contingency scenario. The simulation results are analyzed using the Trajectory Violation Integral (TVI) to quantify voltage deviations ($\Delta V_{i}(t)$) at the system's critical buses. If the buses exhibit no significant voltage deviation, this indicates that the compensation determined in Step 1 is sufficient to maintain voltage stability under contingency conditions. However, if voltage deviations are detected at any bus, a Q-V analysis is conducted for that particular bus to calculate the additional reactive power demand ($\Delta Q_{i}$) required. The reactive power compensation equipment sizing from Step 1 is then updated accordingly, and the RMS simulation is repeated to verify that voltage deviations have been adequately compensated. The additional required reactive power demand are then given as
\begin{equation}
\mathbf{\Delta Q}_{\mathrm{short}} =
\begin{bmatrix}
\Delta Q_{\mathrm{short},1}\\
\Delta Q_{\mathrm{short},2}\\
\vdots\\
\Delta Q_{\mathrm{short},N}
\end{bmatrix}
\end{equation}

A final QDS simulation is performed to ensure that the voltages remain within the pre-defined voltage band after updating the reactive power equipments in the second step with additional reactive power.

\subsection{Assessment and Simulation Tools}
Managing reactive power is regarded one of the most challenging tasks in power system planning and operation, as it involves a large number of parameters and uncertainties. In the context of increasing network loading, maintaining voltage stability has become of increasing significance. Due to the inherently nonlinear and varied nature of power systems, direct analytical methods are often not feasible. This underscores the significance of simulation as a tool for system analysis and design. In the field of power systems, there are three main types of simulations, namely dynamic, time-series, and steady-state simulations.

Dynamic simulations, such as RMS- and EMT-simulations, are based on the differential-algebraic equations (DAE) of the system and time is handled implicitly. This type of simulation captures the fast dynamics of a system, usually covering a range of a few seconds to several minutes. Time series simulations consider the algebraic equations (AE) and address time explicitly. Unlike dynamic simulation, time series simulations ignore the fast-changing dynamics of the system, solving the AE under the assumption of steady state conditions. The solution changes only if the parameters related to slow dynamics change. An example for time series simulation is quasi-dynamic simulation. Steady-state simulations, such as power flow simulations, solve  the AE only at a specific point in time.

In this work, we consider RMS, quasi-dynamic, and Q-V simulations. A Q-V analysis (also known as Q-V-curve) is a tool used in power system studies to analyze voltage stability \cite{Kundur_Power_1994}. It illustrates the relationship between the reactive power $Q$ injected or absorbed at a specific bus and the resulting voltage $V$ at that bus. The vertical axis represents the reactive power, while the horizontal axis represents the bus voltage in per unit (p.u.). Reference \cite{overbye_q-v_1994} provides detailed information on Q-V-curve interpretations.

\begin{figure}[t]
  \centering
  \includegraphics[width=\columnwidth]{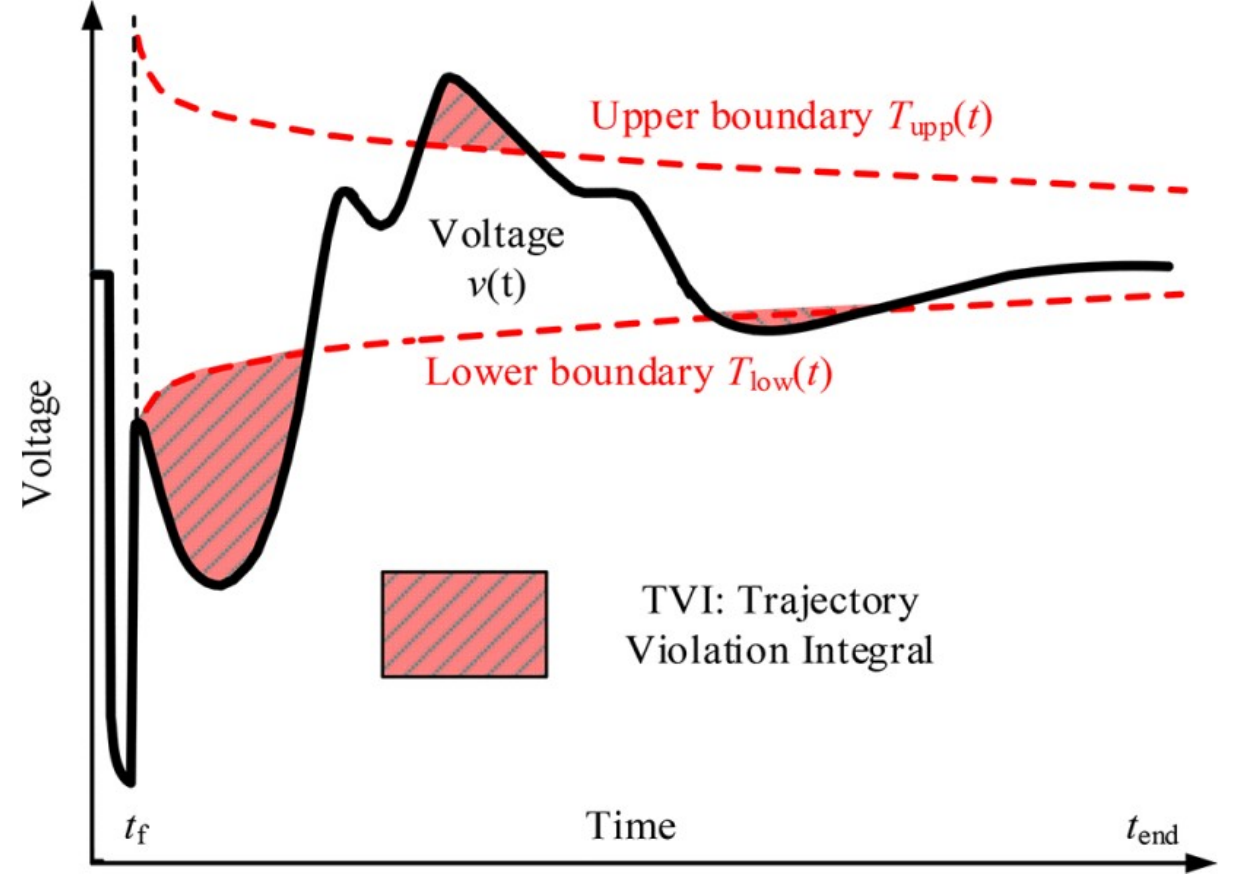}
  \caption{Illustrative graph showing TVI and its boundaries \cite{wildenhues_optimal_2015}}
  \label{fig_TVI}
\end{figure}

In addition, the Trajectory Violation Integral (TVI) is used to assess the dynamic voltage behavior, i.\,e.\ voltage recovery after a fault. TVI is a quantitative metric for the short-term dynamic performance of a power system after a disturbance \cite{dynamic_performance_committee_2024, wildenhues_optimal_2015}. It measures the severity of post-disturbance voltage deviations by calculating the area where the voltage trajectory leaves a predefined, exponentially recovering interval, see Figure \ref{fig_TVI}. As such, TVI provides a comprehensive measure that considers both the magnitude and duration of voltage violations.

The lower boundary of the TVI can be calculated as
\begin{equation}
\label{eqn:T_low}
T_{\text{low}}(t) =
\frac{\left(\frac{t}{t_{\text{end}}} \cdot e^{\frac{t}{t_{\text{end}}}}\right)^{\beta}}{e^{\beta}}
\cdot v_{\text{st}},
\end{equation}
where the parameter $\beta$ determines the rate at which the boundaries decay towards the steady-state voltage $v_\text{st}$ and $t_\text{end}$ is the simulation time. Smaller $\beta$ correspond to more strict voltage requirements. Common values are $\beta \in \left[0, 0.1\right]$. The upper boundary of the TVI is given by
\begin{equation}
\label{eqn:T_upp}
T_\text{upp}(t) = 2 - T_\text{low}(t).
\end{equation}
The TVI is then given by

\begin{equation}
\text{TVI} = \int_{\text{tend}}^{\text{tf}} v(t) \, dt
\end{equation}

\begin{equation}
v(t) =
\begin{cases}
T_{\text{low}}(t) - v(t) & \text{if } v(t) < T_{\text{low}}(t) \\
v(t) - T_{\text{upp}}(t) & \text{if } v(t) > T_{\text{low}}(t) \\
0 & \text{otherwise}
\end{cases}
\end{equation}

corresponding to the indicated area in Fig.~\ref{fig_TVI}. The TVI concept is discussed in greater detail in \cite{wildenhues_optimal_2015}, including its practical application.

\section{Study case}
\label{sec:study_case}

In this Section, the effectiveness of the proposed methodology is verified by power systems simulations. The simulations have been performed with the software tool PowerFactory \cite{noauthor_powerfactory_nodate}.

\subsection{Test Network}
A slightly modified IEEE 39 bus system is used as a testbed for the proposed methodology. The system features standard IEEE dynamic models for synchronous machines, including Automatic Voltage Regulators (AVR), Power System Stabilizers (PSS) and governors \cite{demetriou_dynamic_2017}. Additionally, time series data for loads and synchronous machines are incorporated into the model \cite{li_creation_2021}. These time series data are entirely synthetic and do not represent any real scenarios. The system contains 10 synchronous generators, 19 loads and 34 transmission lines \cite{athay_practical_1979}.

\subsection{Simulations and Results}

\subsubsection*{Step 1: Reactive power demand for long-term voltage stability}

\begin{figure}%[htbp]
  \centering
  \includegraphics[width=\columnwidth]{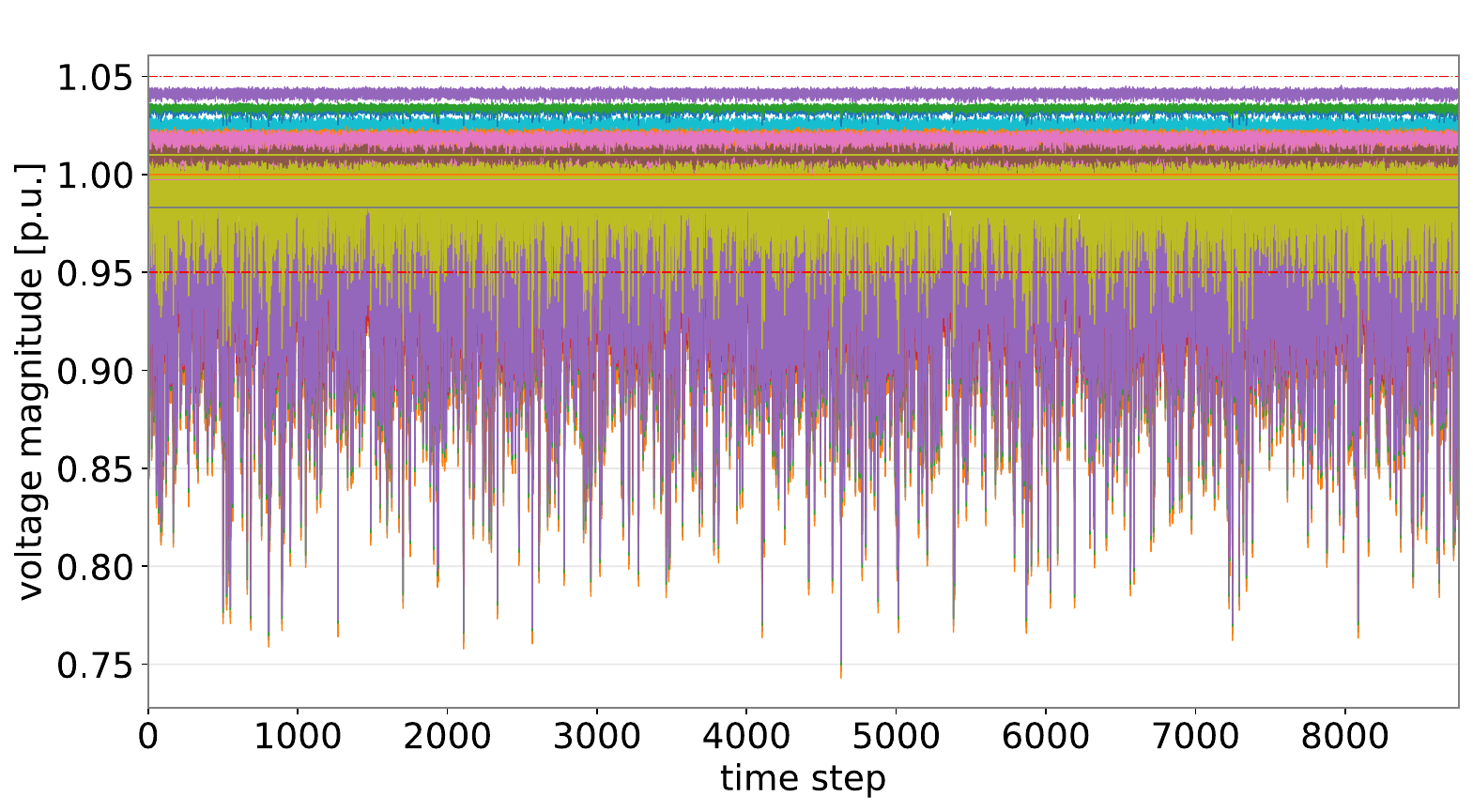}
  \caption{Voltage magnitudes of all buses in the 39-bus system in iteration 1}
  \label{fig_QDS_iter1}
\end{figure}

As mentioned earlier in Section~\ref{sec:methodology}, a QDS is conducted over a single year period (8760 hours) and the voltage magnitude is computed for each bus in the transmission system, as displayed in Figure~\ref{fig_QDS_iter1}. The voltage band is chosen to $\mathcal{V}=[\qty{0.95}{p.u.}, \qty{1.05}{p.u.}]$. It is shown as two horizontal red lines. In normal operation conditions, the voltages should remain within the voltage band.

\begin{figure}%[htbp]
  \centering
  \includegraphics[width=\columnwidth]{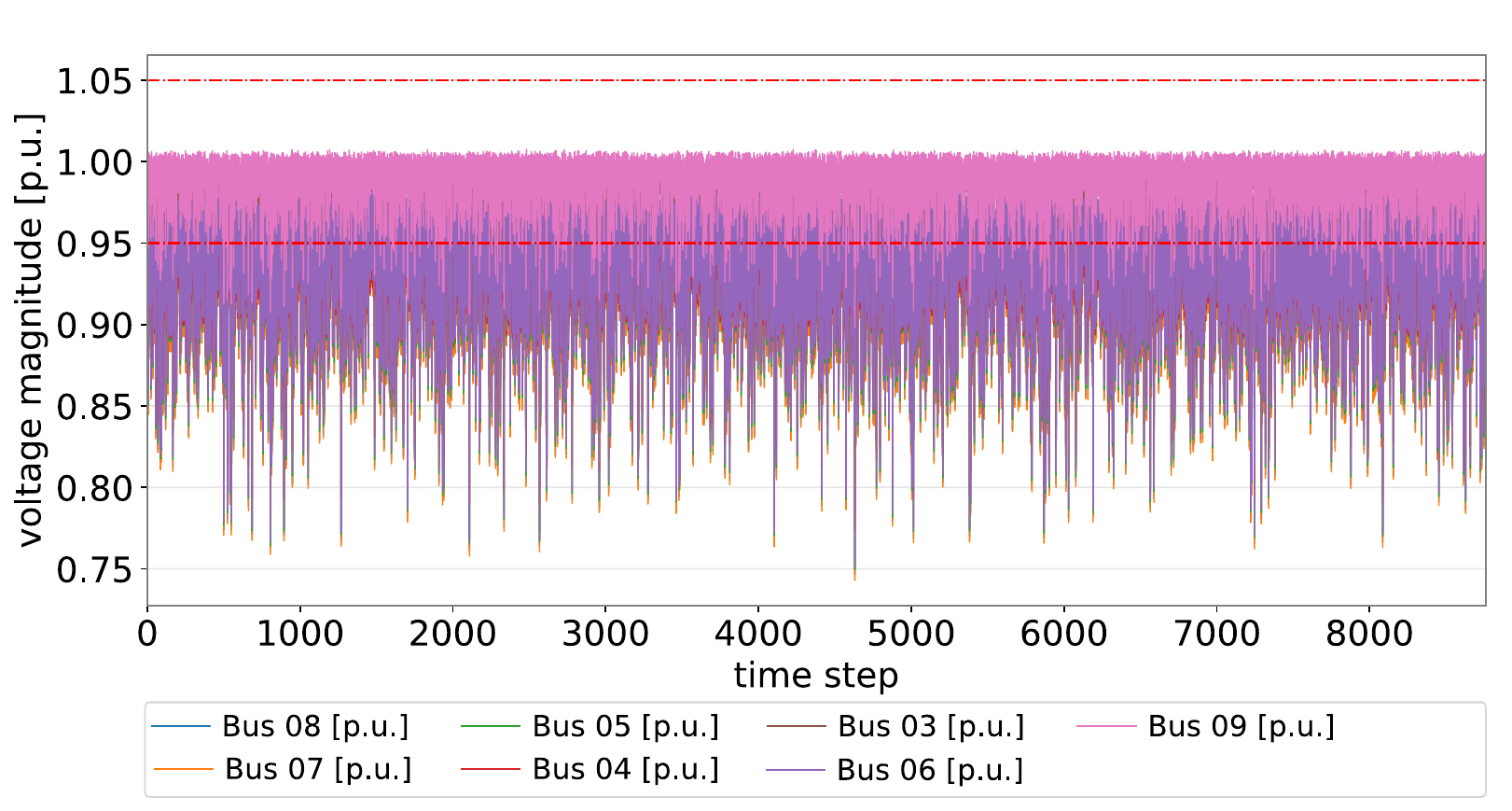}
  \caption{Voltage magnitudes of the buses that violate the voltage band in iteration 1}
  \label{fig_QDS_iter1.1}
\end{figure}

%\vspace{-8pt}
Figure~\ref{fig_QDS_iter1.1} shows only the buses at which voltage violations occur in iteration 1. It can be observed that the voltage profile of multiple buses violates the lower voltage limit. The most critical time step (hour of the year) is 4632, at which the largest voltage violation occurs at 0.743 p.u. 

%\vspace{-15pt}

\begin{figure}%[htbp]
  \centering
  \includegraphics[width=\columnwidth]{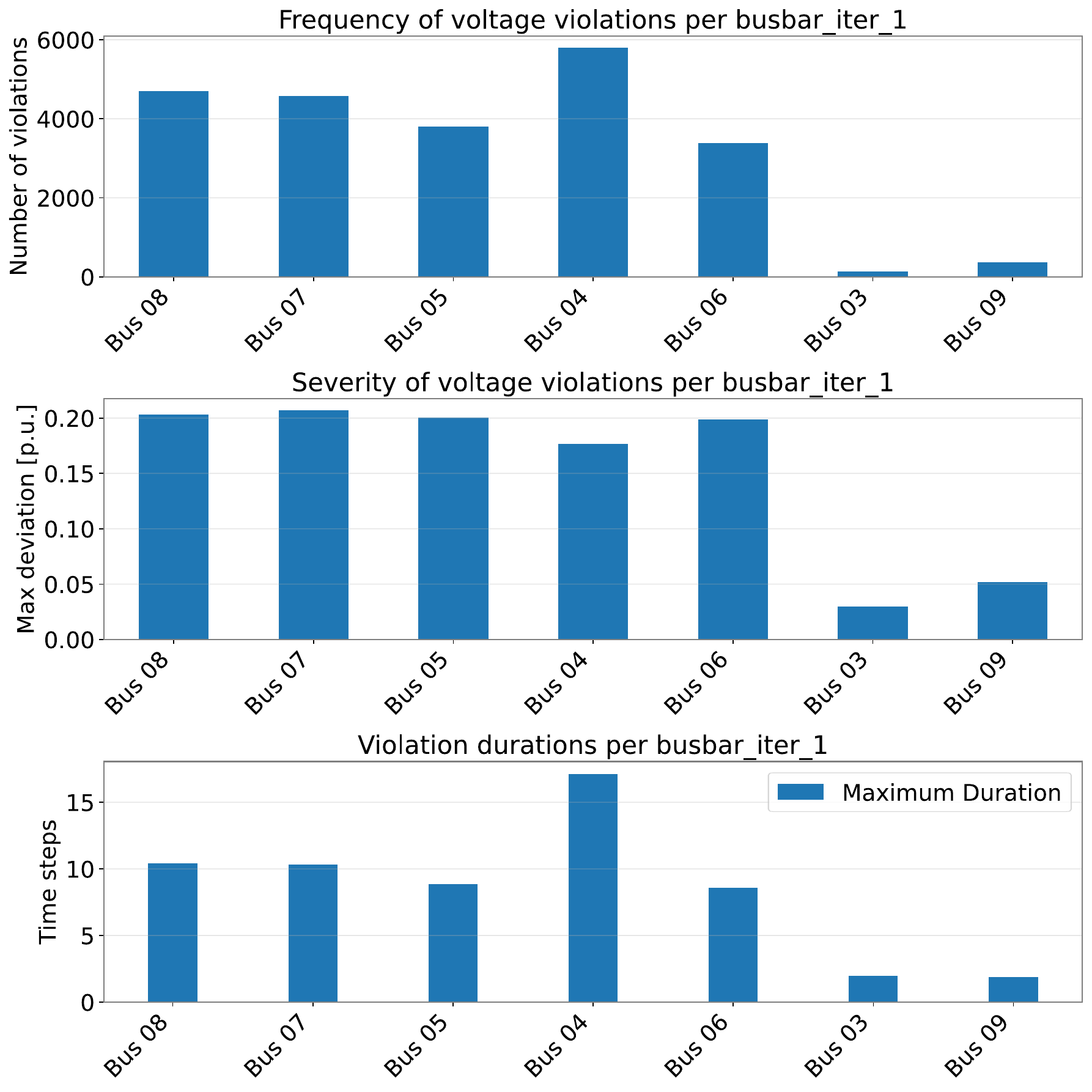}
  \caption{Analysis of simulation results using generic metrics in iteration 1}
  \label{fig_stats_iter1}
\end{figure}

%\vspace{-8pt}
The results are analyzed further using the generic metrics as mentioned in Section~\ref{sec:methodology} to evaluate the bus. It can be concluded that bus 04 is the most critical in the system with a total number of 5800 violations, see Figure~\ref{fig_stats_iter1}.

%\vspace{-15pt}

\begin{figure}%[htbp]
  \centering
  \includegraphics[width=\columnwidth]{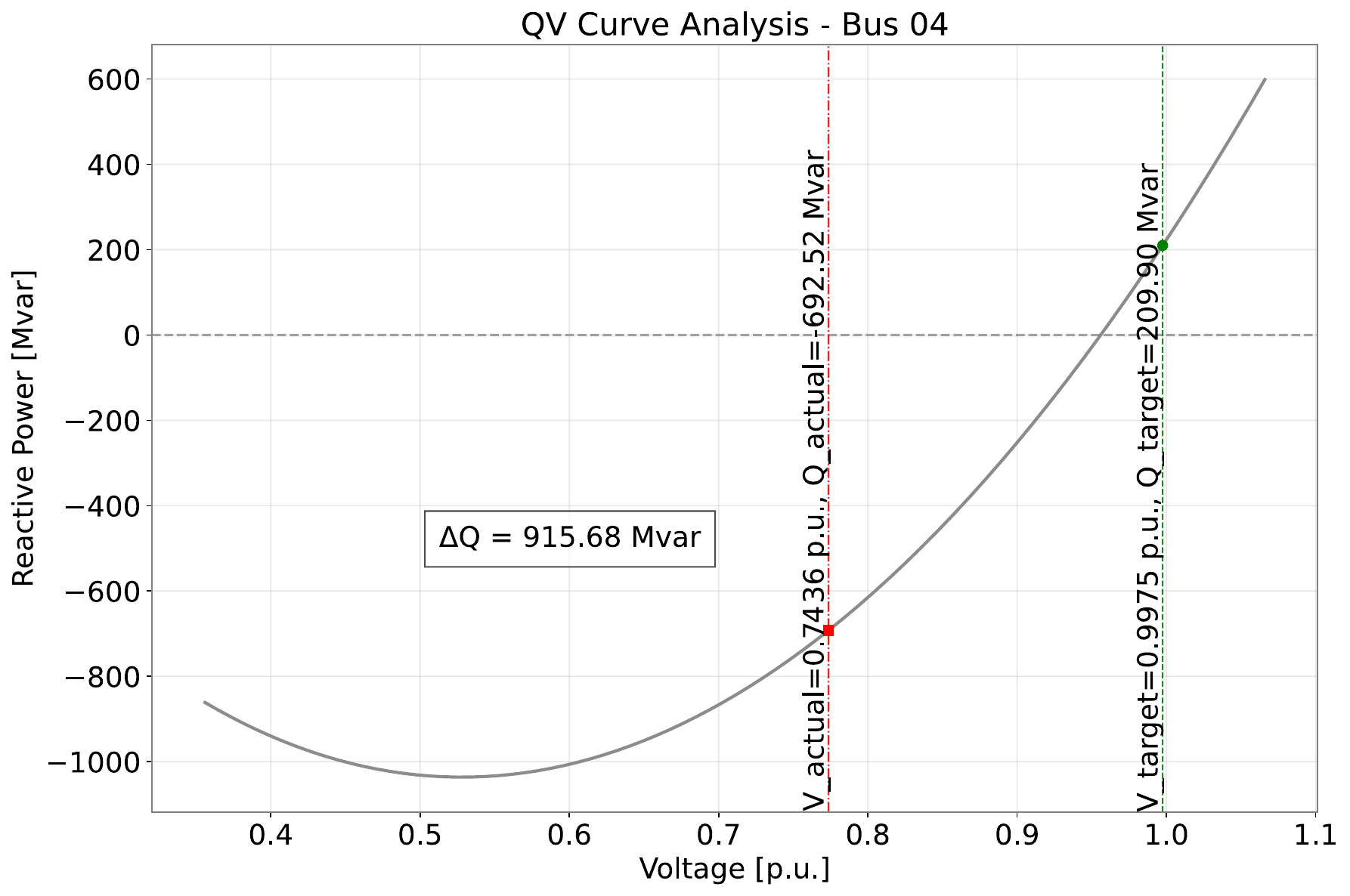}
  \caption{Q-V Curve for Bus 04 in iteration 1}
  \label{fig_Q-V-QDS_iter1}
\end{figure}

The next step involves calculating the required reactive power $\Delta Q_{long,4}$ to ensure that the voltage at the most critical bus, bus 04, is within the allowable voltage band. A Q-V simulation is performed for this particular bus and the Q-V curve is plotted as shown in Figure~\ref{fig_Q-V-QDS_iter1}. The target voltage is assumed to be 5\% above the minimum voltage limit. This is equivalent to 0.9975 p.u.

The required reactive power $\Delta Q_{long,04}$ to bring the voltage to the target voltage within the allowable voltage band is \qty{916}{Mvar}. Reactive power compensation equipment is connected to this particular bus, Bus 04, and another QDS, namely iteration 2, is performed to verify the effectiveness of the compensation equipment.

\begin{figure}%[htbp]
  %\centering
  \includegraphics[width=\columnwidth]{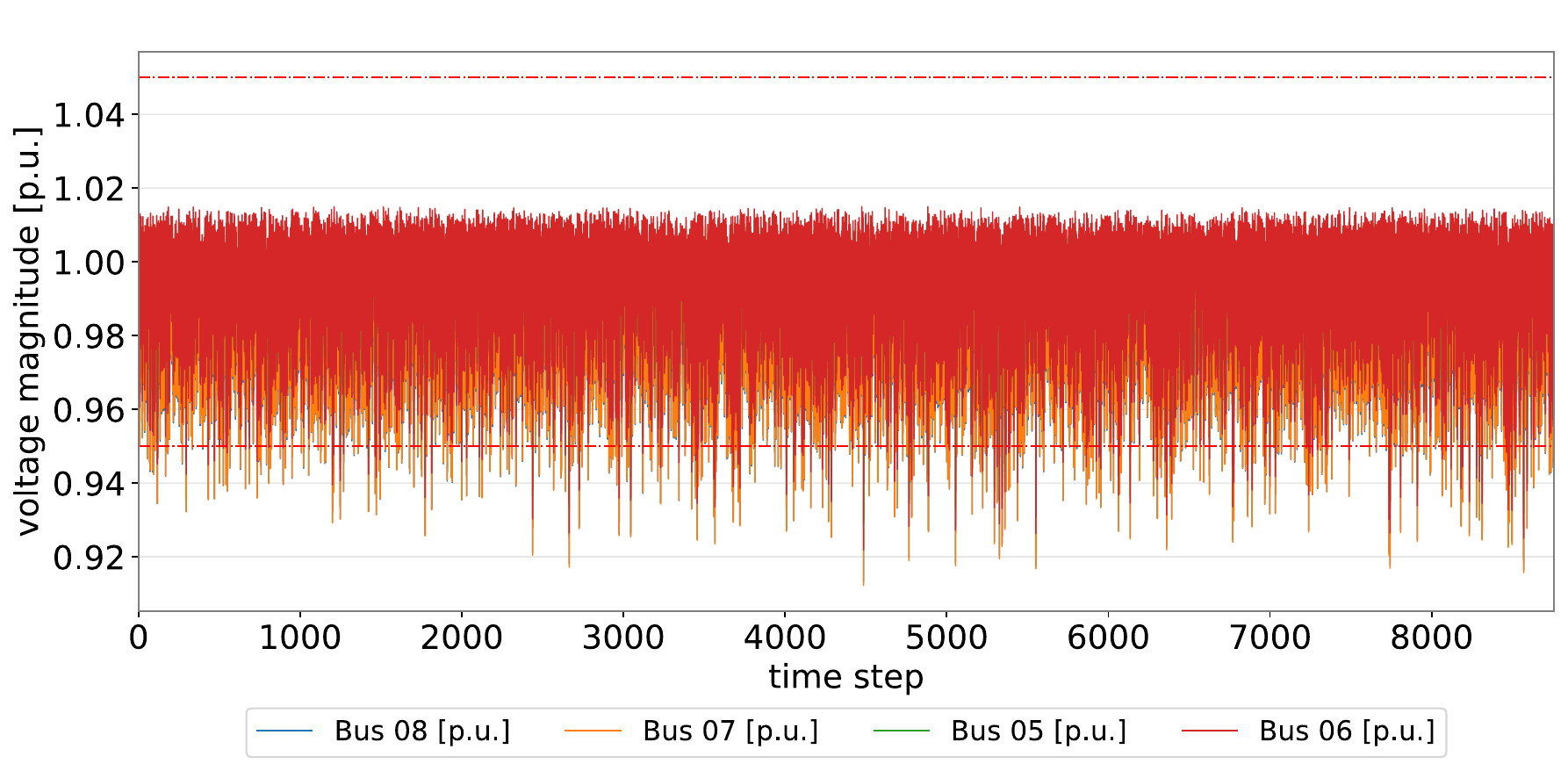}
  \caption{Voltage magnitudes of the buses that violate the voltage band in iteration 2}
  \label{fig_QDS_iter2}
\end{figure}

Figure \ref{fig_QDS_iter2} displays the voltage profiles of the violating buses after the compensation equipment was added to Bus 04. It is evident that, even though some voltages are still critical, all voltages have been improved compared to iteration 1. Another iteration is necessary to further eliminate remaining critical voltages.

\begin{figure}%[htbp]
  \centering
  \includegraphics[width=\columnwidth]{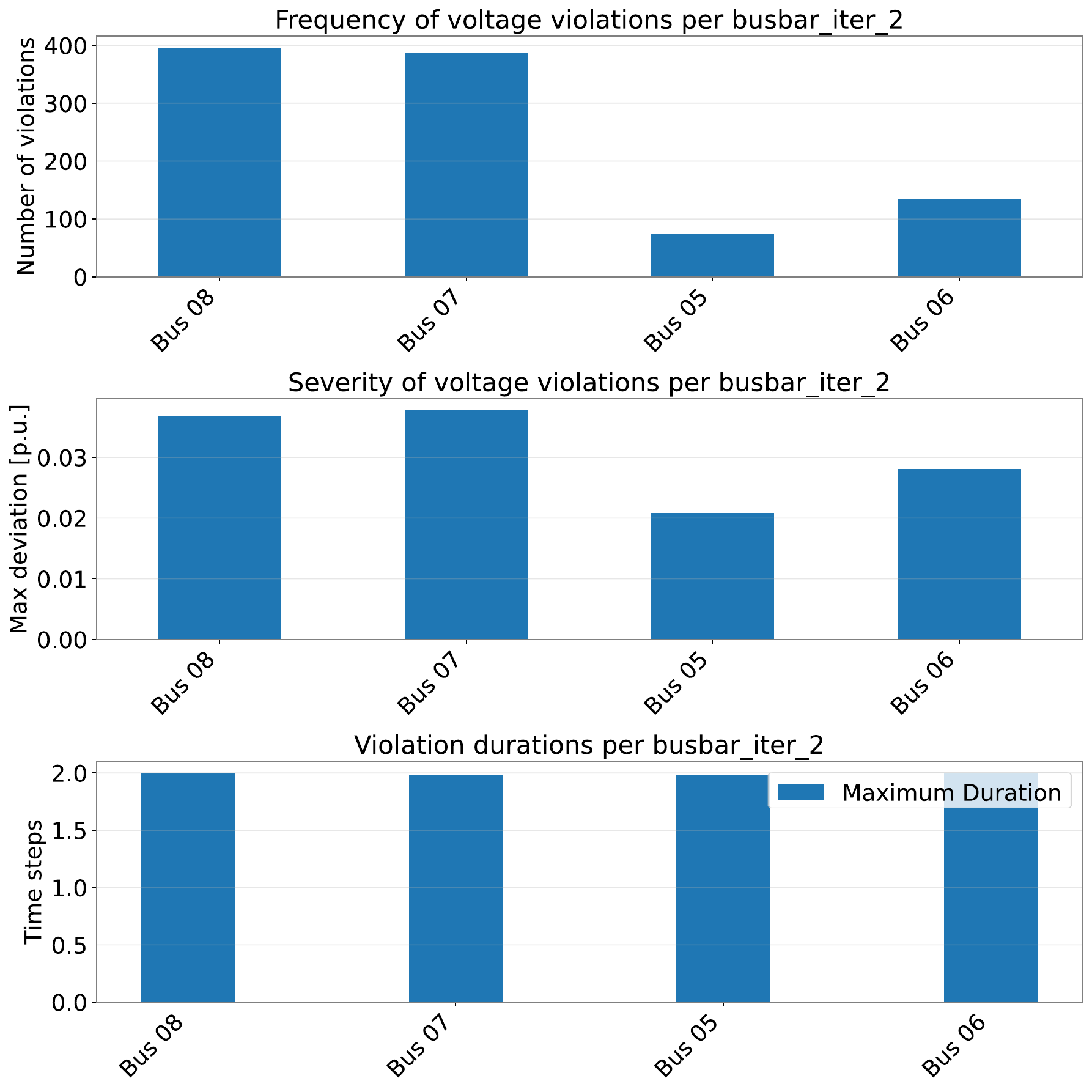}
  \caption{Analysis of simulation results using generic metrics in iteration 2}
  \label{fig_stats_iter2}
\end{figure}

The results presented in Figure \ref{fig_stats_iter2} show that the voltages on buses Bus 04, Bus 03, and Bus 09 are not experiencing any violations. Thus, they remain in the allowed voltage band after the second iteration. Additionally, it can be concluded that Bus 08 is the next most critical in the system with a total number of 400 violations, see Figure \ref{fig_stats_iter2}.

\begin{figure}%[htbp]
  \centering
  \includegraphics[width=\columnwidth]{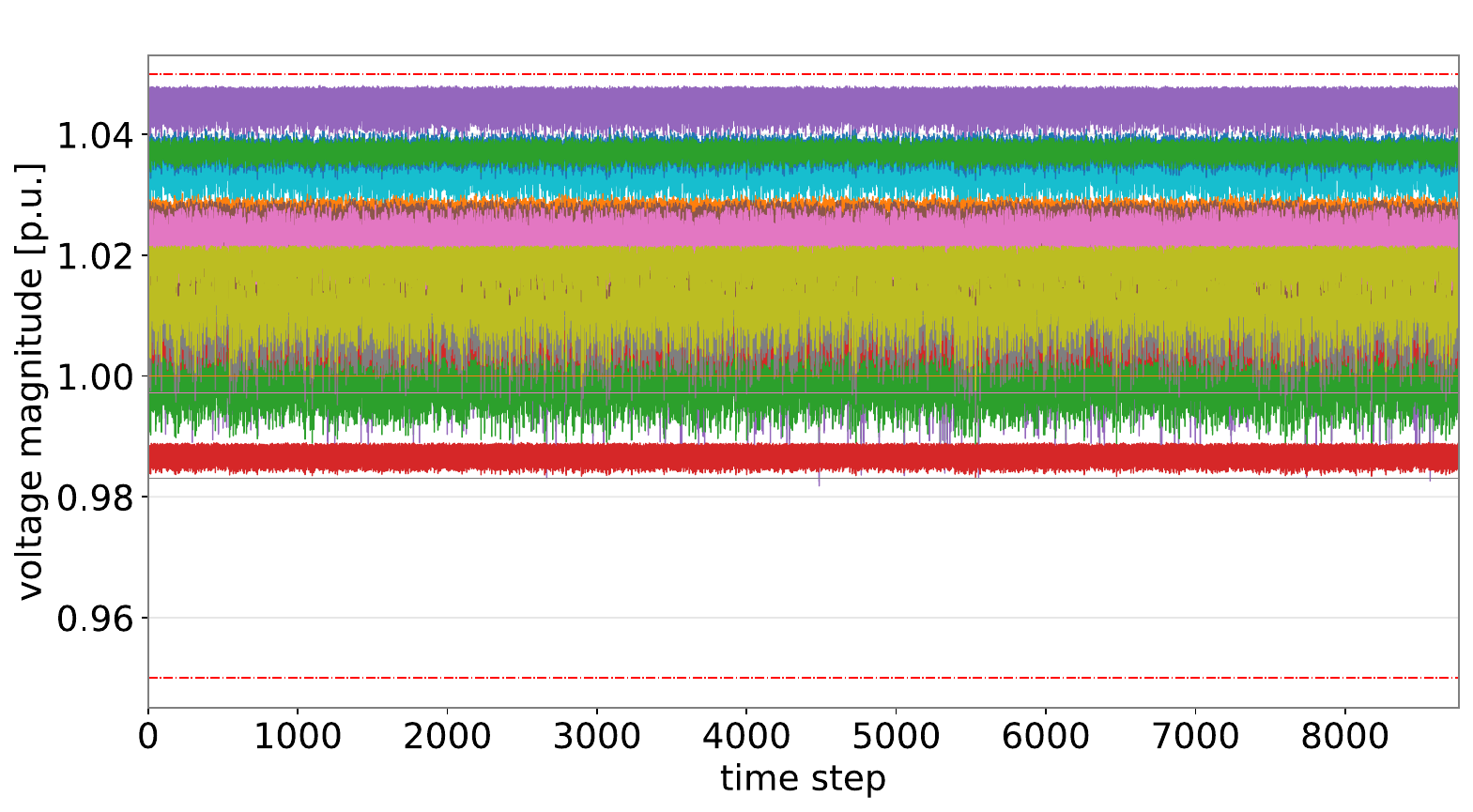}
  \caption{Voltage magnitudes of all buses in the transmission system in iteration 3}
  \label{fig_QDS_iter3}
\end{figure}

Similar to Bus 04 in the first iteration, a Q-V analysis was conducted and the required reactive power was calculated. A compensation equipment was added on Bus 08 with $\Delta Q_{long,08} = \qty{606}{Mvar}$ and another QDS simulation was conducted, namely iteration 3. The results in Figure \ref{fig_QDS_iter3} show that all voltages in the system are now operating within the pre-defined limits and no further voltage violations are observed.

In summary, a total of three iterations were needed to calculate the reactive power demand $Q_\mathrm{Long} = \qty{1522}{Mvar}$ for long-term voltage stability. The compensation equipments were added to the most critical buses, Bus 04 and Bus 08, to bring the voltages within the allowed operating voltage band. It is worth noting that the calculated reactive power demand had a system-wide impact improving the overall voltage profiles, rather than only the voltage profiles on the buses where compensation equipment is installed.

\subsubsection*{Step 2: Reactive power demand for short-term voltage stability}

The results obtained from Step 1 are used as input for Step 2, where only the critical time steps (here: 4632) at which minimum voltages occur, are considered for further analysis. This time step represents the worst-case operating condition, characterized by critically low voltage levels. For the TVI analysis a strict voltage requirement with $\beta$ assumed as 0.015.

\begin{figure}%[!htbp]
  \centering
  \includegraphics[width=\columnwidth]{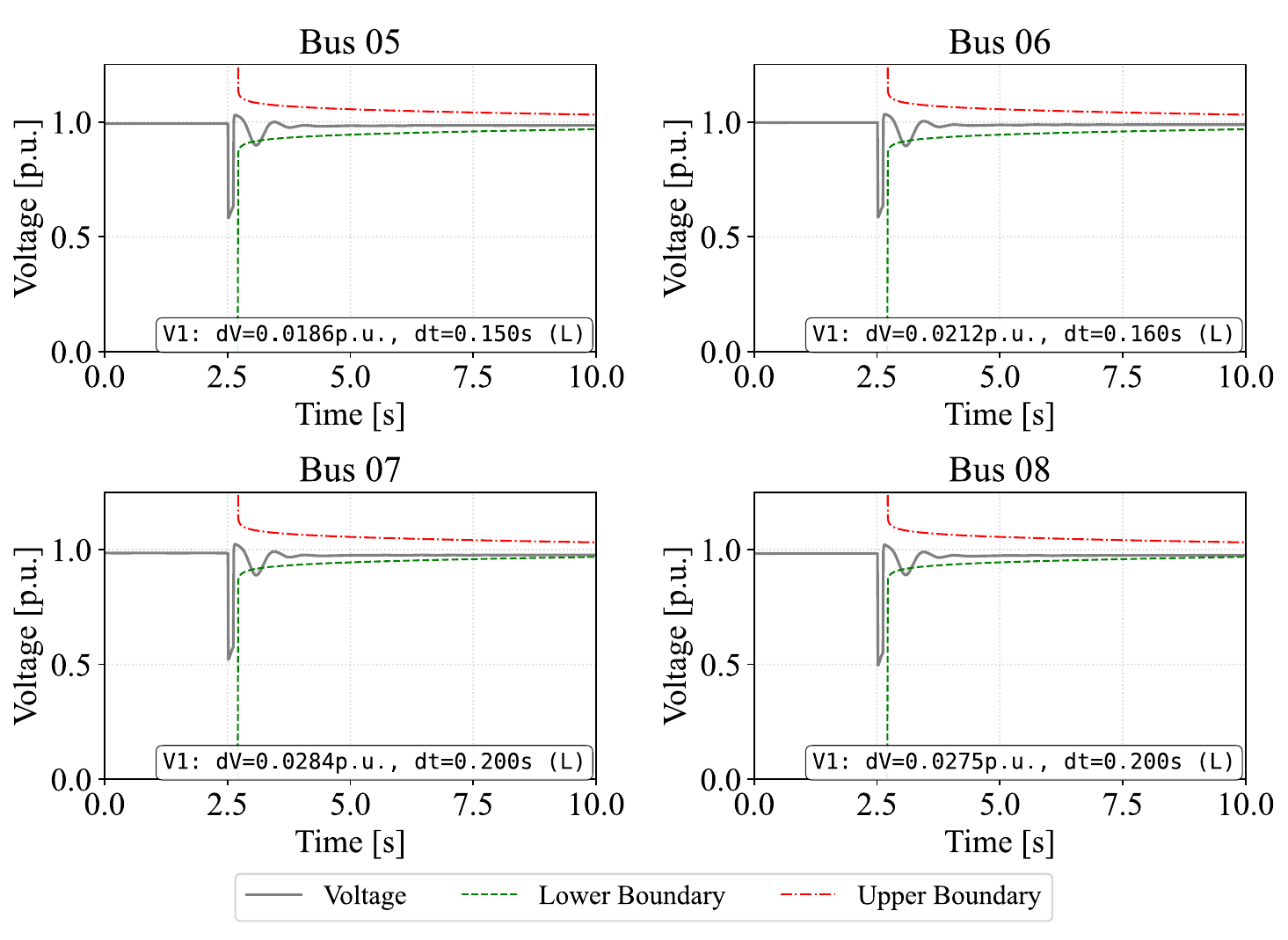}
  \caption{TVI analysis and voltage curves of all buses in iteration 1}
  \label{fig_TVI_iter1}
\end{figure}

A contingency list was defined including a short circuit fault on line 09-39 at $t=\qty{2.5}{s}$ with a fault clearing time of 100 ms combined with a generator event in which G 03 went out of service at $t=\qty{2.66}{s}$. These two events have caused a large disturbance in the system with a noticeable voltage drop following the short-circuit recovery, see \ref{fig_TVI_iter1}.

\begin{figure}%[htbp]
  \centering
  \includegraphics[width=\columnwidth]{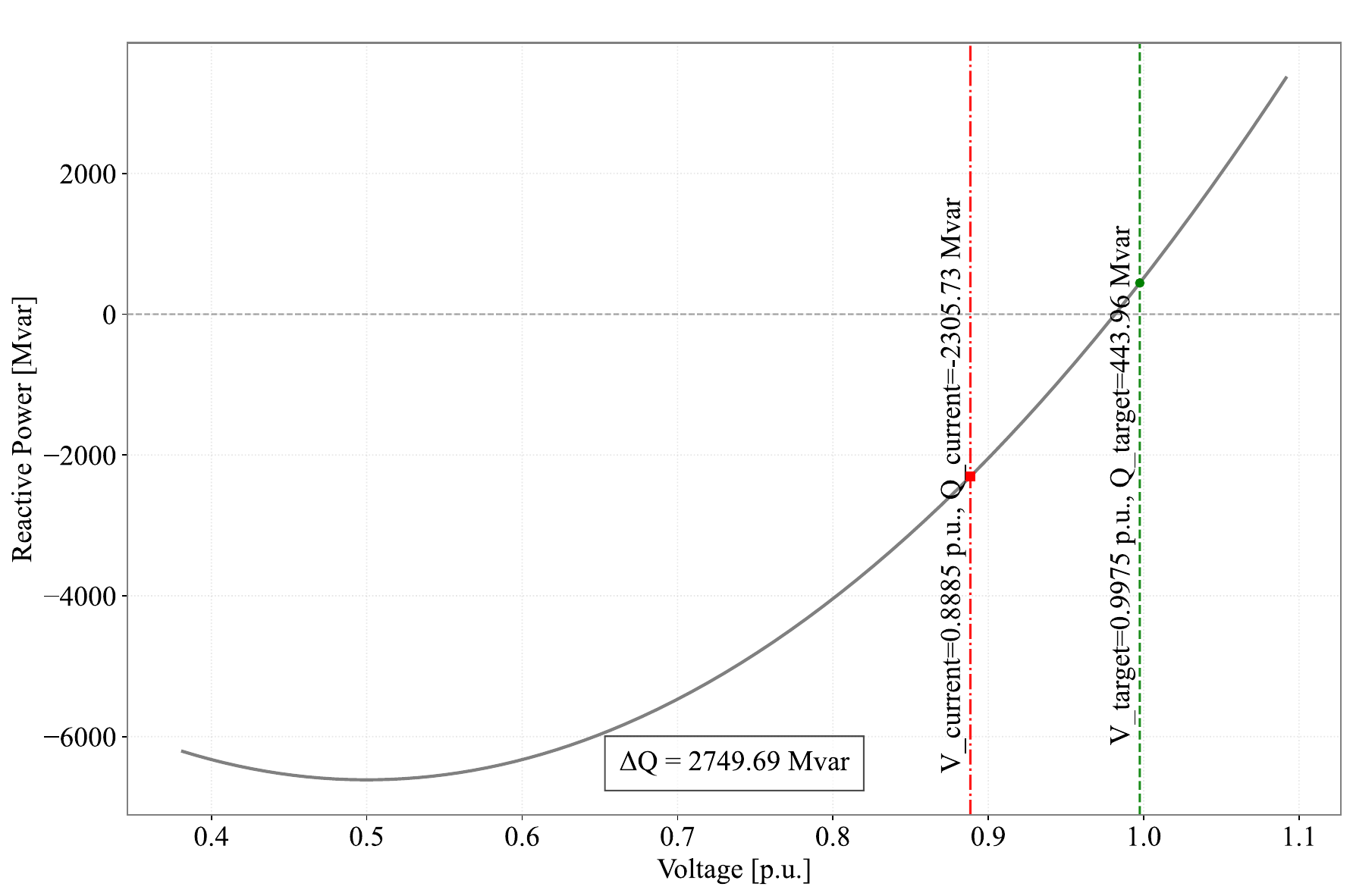}
  \caption{Q-V Analysis for Bus 07 after iteration 1}
  \label{fig_Q-V-RMS_iter1}
\end{figure}

The results were then analyzed using the TVI to determine whether a voltage deviation above or below the TVI-limits exists. It is evident that the lower TVI-limits are violated at multiple buses. As shown in Figure \ref{fig_TVI_iter1}, the bus with the largest deviations ($dV = 0.0284$ p.u.) is Bus 07.

A Q-V analysis for Bus 07 was conducted and the reactive power $\Delta Q_\mathrm{short,07}$ required to keep the voltage within the TVI-limits was calculated. According to Figure~\ref{fig_Q-V-RMS_iter1}, a compensation device providing $\Delta Q_\mathrm{short,07}=\qty{2750}{Mvar}$ is required to keep the voltage within the TVI-limits.

\begin{figure}
  \centering
  \includegraphics[width=\columnwidth]{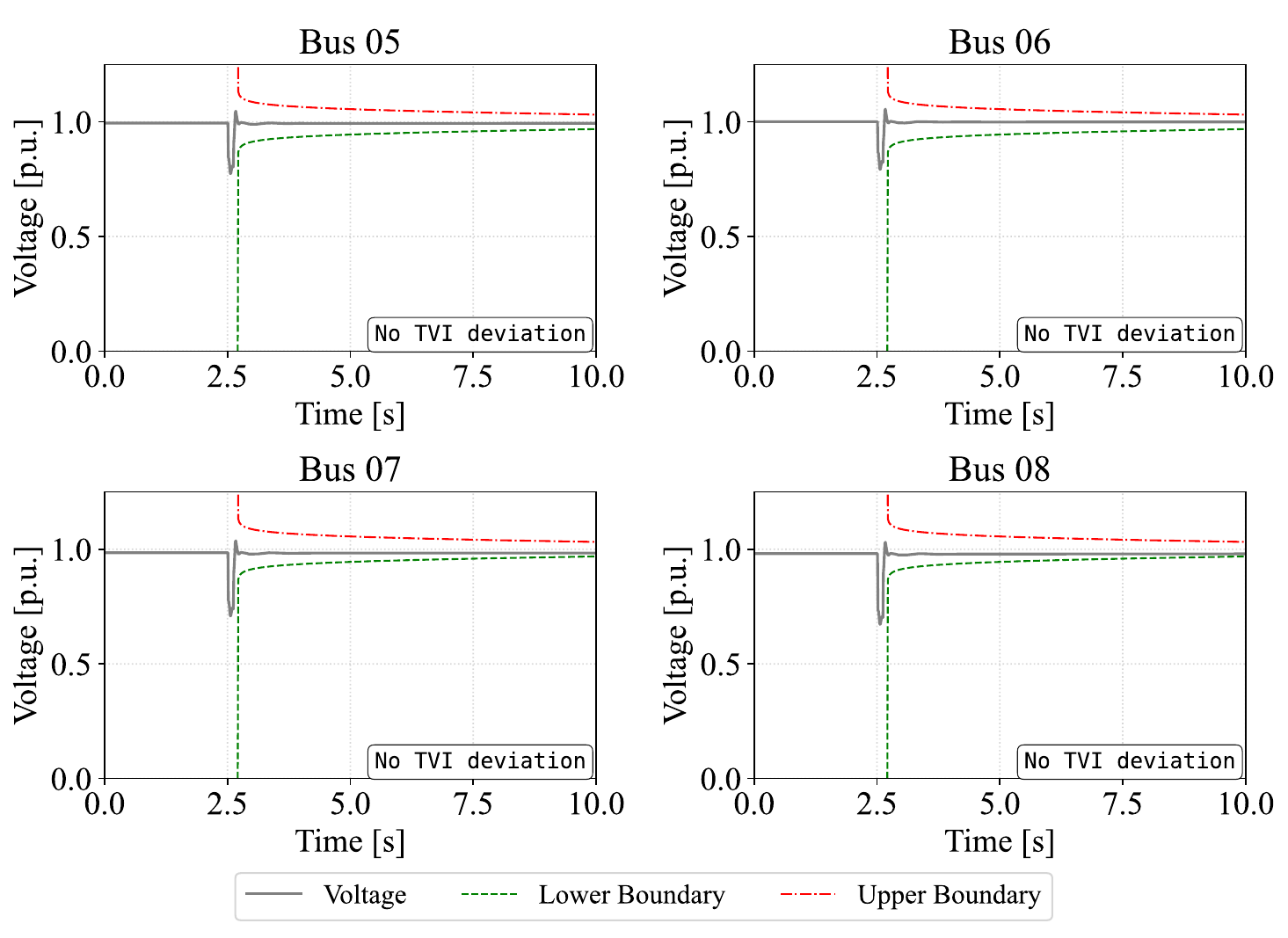}
  \caption{TVI analysis and voltage curves of all buses in iteration 2}
  \label{fig_TVI_iter2}
\end{figure}

A dynamic compensation device was added to Bus 07 with $\Delta Q_{short,07}=\qty{2750}{Mvar}$ and the RMS simulation was repeated. Figure \ref{fig_TVI_iter2} displays the resulting voltage curves. It can be concluded that the voltage violations on all buses from iteration 1 were addressed and eliminated.

\begin{figure}%[!htbp]
  \centering
  \includegraphics[width=\columnwidth]{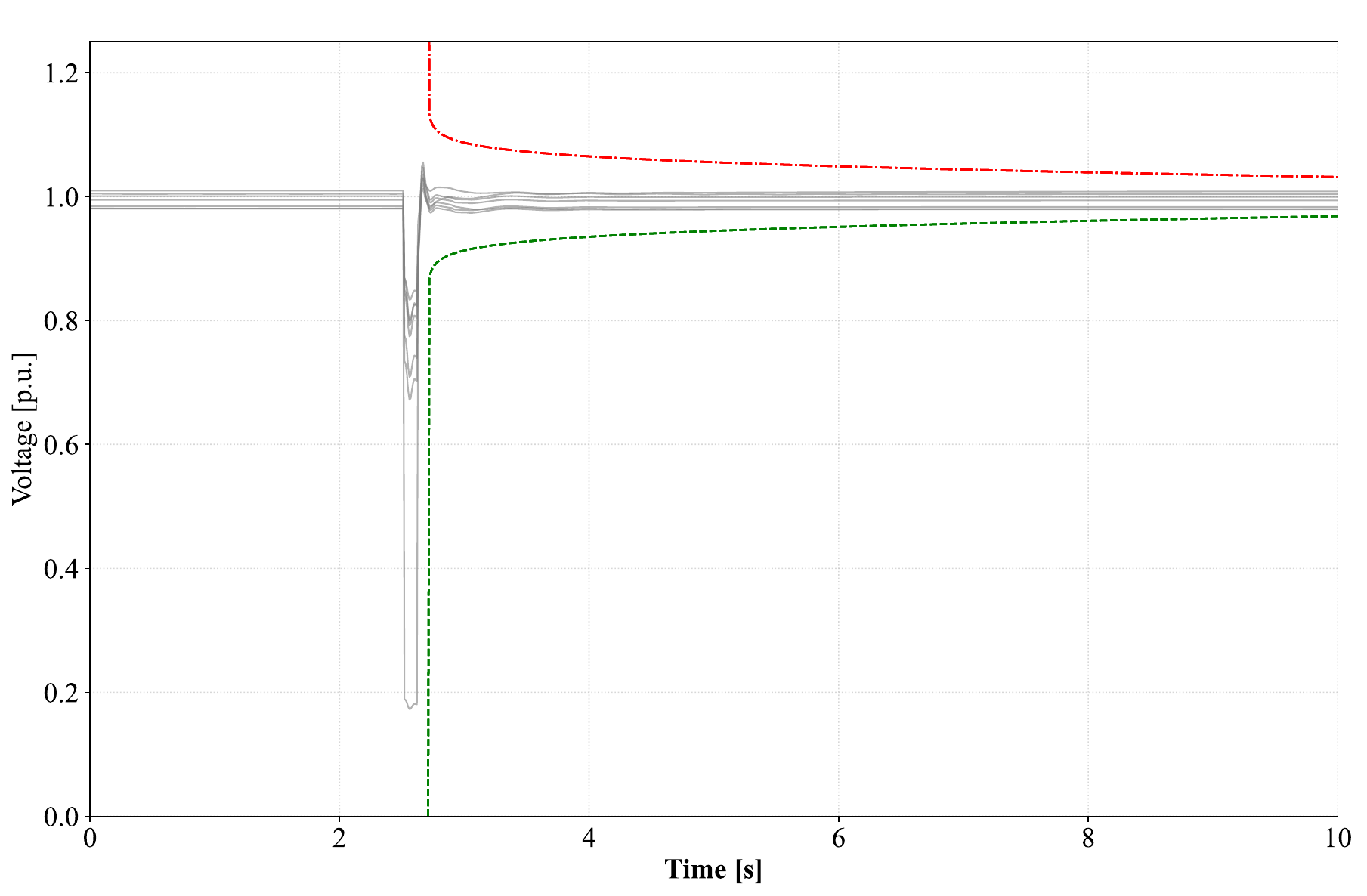}
  \caption{TVI analysis and voltage curves of all buses – after compensation in step 2}
  \label{fig_TVI_iter3}
\end{figure}

Figure \ref{fig_TVI_iter3} shows the final results of the RMS simulations after adding the dynamic compensation equipment in iteration 1. The voltage curves on all buses of the system are now operating within the TVI-boundaries and show no voltage deviations.

In summary, two iterations were required to ensure short-term voltage stability.%, leading to a total calculated reactive power demand of $Q_\mathrm{Short} = \qty{727}{Mvar}$.

%\newpage

\section{Conclusion and Outlook}
\label{sec:conclusion}

This paper presents a novel methodology for calculating reactive power demand in transmission networks through a comprehensive two-step simulation framework. Unlike existing methods that rely on single simulations or optimization-based approaches, the proposed methodology integrates Quasi-Dynamic Simulation (QDS) and RMS simulations with Q-V curve analysis in an iterative sequential process to directly calculate the actual reactive power requirements. The approach uniquely combines both long-term and short-term voltage stability assessments, thereby capturing both slow variations in operating conditions and fast dynamic phenomena that may otherwise be overlooked when only one timescale is considered. By analyzing a full annual period (8760 hours) using multi-criteria metrics including number, severity, and duration of voltage violations, the methodology provides a richer characterization of system performance. It can be straightforwardly extended to incorporate alternative indices such as Fast Voltage Stability Index (FVSI), Line Stability Factor (LQP), or Voltage Stability Margin (VSM) according to the needs of the study. The framework has been tested on four network models of different sizes, including a system with more than 50 nodes, demonstrating its scalability and applicability to both transmission and distribution networks. In general, this methodology provides a systematic and practical framework that can be applied to any network, offering the flexibility for engineers to incorporate their own assessment tools, performance indices, and evaluation criteria. On the other hand, the proposed methodology calculates the reactive power demand specifically for those buses that suffer from voltage violations. This could lead to an overestimation of the actual reactive power demand. Moreover, in real networks it is not feasible to install a compensation equipment at each critical bus that experiences voltage deviations due to space limitations and regulatory constraints.

Looking ahead, future work will focus on applying the proposed approach to large-scale power systems with more than 100,000 nodes to further validate its computational efficiency and robustness for realistic grid sizes. In addition, dedicated voltage stability metrics will be developed and integrated into the workflow to automatically identify critical buses and critical hours, enabling more targeted mitigation measures and more effective planning. Moreover, the authors are investigating methods to optimize the calculated reactive power demand with respect to sizing and placement.

\section*{Acknowledgments}

This work was conducted as part of the research project \emph{SysStab2030}, financed by the German Ministry for Economic Affairs and Energy with grant number 03EI6122K.

The authors express their gratitude to the \emph{Voltage stability} working group of the four German transmission network operators for their valuable feedback and insights during this research. In addition, the authors thank Sebastian Wildenhaus for providing detailed information about the TVI analysis.

The authors declare no conflicts of interest.

\printbibliography

@article{wildenhues_optimal_2015,
	title = {Optimal Allocation and Sizing of Dynamic Var Sources Using Heuristic Optimization},
	volume = {30},
	issn = {0885-8950, 1558-0679},
	doi = {10.1109/TPWRS.2014.2361153},
	pages = {2538--2546},
	number = {5},
	journaltitle = {{IEEE} Transactions on Power Systems},
	shortjournal = {{IEEE} Trans. Power Syst.},
	author = {Wildenhues, Sebastian and Rueda, Jose L. and Erlich, Istvan},
	date = {2015-09},
	langid = {english},

}

@article{boricic_comprehensive_2021,
	title = {Comprehensive Review of Short-Term Voltage Stability Evaluation Methods in Modern Power Systems},
	volume = {14},
	issn = {1996-1073},
	doi = {10.3390/en14144076},
	pages = {4076},
	number = {14},
	journaltitle = {Energies},
	shortjournal = {Energies},
	author = {Boričić, Aleksandar and Torres, José L. Rueda and Popov, Marjan},
	date = {2021-07-06},
	langid = {english},

}

@inproceedings{fang_03_reactive_2014,
	location = {Chicago, {IL}, {USA}},
	title = {Reactive power planning considering high penetration of wind energy},
	isbn = {978-1-4799-3656-4},
	doi = {10.1109/TDC.2014.6863471},
	eventtitle = {2014 {IEEE}/{PES} Transmission \& Distribution Conference \& Exposition (T\&D)},
	pages = {1--5},
	booktitle = {2014 {IEEE} {PES} T\&D Conference and Exposition},
	publisher = {{IEEE}},
	author = {Fang, Xin and Li, Fangxing and Xu, Yan},
	date = {2014-04},
	langid = {english},

}

@inproceedings{pranto_04_reactive_2021,
	location = {Jakarta, Indonesia},
	title = {Reactive Power Planning of Power System Using Trajectory Sensitivity Index},
	isbn = {978-1-6654-1641-2},
	doi = {10.1109/ICT-PEP53949.2021.9600933},
	eventtitle = {2021 International Conference on Technology and Policy in Energy and Electric Power ({ICT}-{PEP})},
	pages = {80--84},
	booktitle = {2021 International Conference on Technology and Policy in Energy and Electric Power ({ICT}-{PEP})},
	publisher = {{IEEE}},
	author = {Pranto, Tarik Ahasan and Aziz, Tareq},
	date = {2021-09-29},
	langid = {english},

}

@article{hatziargyriou_definition_2021,
	title = {Definition and Classification of Power System Stability – Revisited \& Extended},
	volume = {36},
	issn = {0885-8950, 1558-0679},
	doi = {10.1109/TPWRS.2020.3041774},
	pages = {3271--3281},
	number = {4},
	journaltitle = {{IEEE} Transactions on Power Systems},
	shortjournal = {{IEEE} Trans. Power Syst.},
	author = {Hatziargyriou, Nikos and Milanovic, Jovica and Rahmann, Claudia and Ajjarapu, Venkataramana and Canizares, Claudio and Erlich, Istvan and Hill, David and Hiskens, Ian and Kamwa, Innocent and Pal, Bikash and Pourbeik, Pouyan and Sanchez-Gasca, Juan and Stankovic, Aleksandar and Van Cutsem, Thierry and Vittal, Vijay and Vournas, Costas},
	date = {2021-07},
	langid = {english},

}

@book{Kundur_Power_1994,
	title = {Power System Stability and Control},
	author = {Prabha Kundur},
    publisher = {McGraw-Hill Education},
	date = {1994},
    isbn = {007035958X, 9780070359581},
	langid = {english},
}

@inproceedings{barot_02_optimal_2007,
	location = {Tampa, {FL}, {USA}},
	title = {Optimal Reactive Power Planning and Compensation Effects on Transmission Loss Components},
	isbn = {978-1-4244-1296-9 978-1-4244-1298-3},
	doi = {10.1109/PES.2007.386161},
	eventtitle = {2007 {IEEE} Power Engineering Society General Meeting},
	pages = {1--7},
	booktitle = {2007 {IEEE} Power Engineering Society General Meeting},
	publisher = {{IEEE}},
	author = {Barot, Hemant and Bhattacharya, Kankar},
	date = {2007-06},
	langid = {english},
	note = {{ISSN}: 1932-5517},

}

@inproceedings{jayasudha_05_sizing_2025,
	location = {Gurugram, India},
	title = {Sizing of a Reactive Power Compensation Device for a 300MW Grid Connected Wind Plant},
	isbn = {979-8-3315-8673-7},
	doi = {10.1109/MRIE66930.2025.11156630},
	eventtitle = {2025 2nd International Conference On Multidisciplinary Research and Innovations in Engineering ({MRIE})},
	pages = {399--404},
	booktitle = {2025 2nd International Conference On Multidisciplinary Research and Innovations in Engineering ({MRIE})},
	publisher = {{IEEE}},
	author = {{Jayasudha} and K H, Vishakh and Nair, Manjula G and Pisini, Jagadeesh Kumar and Thomas, Anil},
	date = {2025-07-30},
	langid = {english},

}

@inproceedings{zhang_06_a_2009,
	location = {Wuhan, China},
	title = {A Novel Reactive Power Compensation Scheme of {UHV} {AC} Transmission Line},
	isbn = {978-1-4244-2486-3},
	doi = {10.1109/APPEEC.2009.4918102},
	eventtitle = {2009 Asia-Pacific Power and Energy Engineering Conference},
	pages = {1--4},
	booktitle = {2009 Asia-Pacific Power and Energy Engineering Conference},
	publisher = {{IEEE}},
	author = {Zhang, Jingchao and Chen, Zhuoya and Gao, Kun and Yan, Anhe},
	date = {2009-03},
	langid = {english},

}

@software{noauthor_powerfactory_nodate,
	title = {{PowerFactory} {DIg} {SILENT}},
	url = {https://www.digsilent.de/en/powerfactory.html},
	version = {2025 {SP}3 (x64)},
	publisher = {{DIg} {SILENT}},
}

@article{li_creation_2021,
	title = {The Creation and Validation of Load Time Series for Synthetic Electric Power Systems},
	volume = {36},
	issn = {0885-8950, 1558-0679},
	doi = {10.1109/TPWRS.2020.3018936},
	pages = {961--969},
	number = {2},
	journaltitle = {{IEEE} Transactions on Power Systems},
	shortjournal = {{IEEE} Trans. Power Syst.},
	author = {Li, Hanyue and Yeo, Ju Hee and Bornsheuer, Ashly L. and Overbye, Thomas J.},
	date = {2021-03},
	langid = {english},

}

@article{athay_practical_1979,
	title = {A Practical Method for the Direct Analysis of Transient Stability},
	volume = {{PAS}-98},
	issn = {0018-9510},
	doi = {10.1109/TPAS.1979.319407},
	pages = {573--584},
	number = {2},
	journaltitle = {{IEEE} Transactions on Power Apparatus and Systems},
	shortjournal = {{IEEE} Trans. on Power Apparatus and Syst.},
	author = {Athay, T. and Podmore, R. and Virmani, S.},
	date = {1979-03},
	langid = {english},

}

@misc{iea_germany,
	title = {Germany 2020 - Energy Policy Review},
	url = {https://iea.blob.core.windows.net/assets/60434f12-7891-4469-b3e4-1e82ff898212/Germany_2020_Energy_Policy_Review.pdf},
	publisher = {International Energy Agency ({IEA})},
	author = {{IEA}},
}

@article{demetriou_dynamic_2017,
	title = {Dynamic {IEEE} Test Systems for Transient Analysis},
	volume = {11},
	issn = {1932-8184, 1937-9234, 2373-7816},
	doi = {10.1109/JSYST.2015.2444893},
	pages = {2108--2117},
	number = {4},
	journaltitle = {{IEEE} Systems Journal},
	shortjournal = {{IEEE} Systems Journal},
	author = {Demetriou, Panayiotis and Asprou, Markos and Quiros-Tortos, Jairo and Kyriakides, Elias},
	date = {2017-12},
	langid = {english},

}

@article{overbye_q-v_1994,
	title = {{Q-V} Curve Interpretations Of Energy Measures For Voltage Security},
	volume = {9},
	doi = {10.1109/59.317593},
	pages = {331 -- 340},
	number = {1},
	author = {Overbye, Thomas J. and Dobson, Ian and L. {DeMarco}, Christopher},
	date = {1994-02},

}

@report{dynamic_performance_committee_2024,
	title = {Evaluation of Voltage Stability Assessment Methodologies in Modern Power Systems with Increased Penetration of Inverter-Based Resources},
	number = {{PES}-{TR}126},
	institution = {The Institute of Electrical and Electronics Engineers, Inc.},
	type = {Technical Report},
	author = {{Technical Committee: Power Systems Dynamic Performance Committee}},
	date = {2024},

}

@inproceedings{li_18_dynamic_2020,
	location = {Shanghai, China},
	title = {Dynamic Reactive Power Allocation Strategy for {AC}/{DC} System Based on Quantum Genetic Algorithm},
	rights = {https://ieeexplore.ieee.org/Xplorehelp/downloads/license-information/{IEEE}.html},
	isbn = {978-1-7281-7687-1},
	eventtitle = {2020 Chinese Automation Congress ({CAC})},
	pages = {1949--1954},
	booktitle = {2020 Chinese Automation Congress ({CAC})},
	publisher = {{IEEE}},
	author = {Li, Yanchun and Bai, Enming and Xu, Jianyuan and Liu, Jicheng and Liu, Shaowu and Ren, Puchun and Yuan, Peng},
	date = {2020-11-06},

}

@article{satyamsetti_19_active_2021,
	title = {ACTIVE COMPENSATION OF REACTIVE POWER VIA {STATCOM} ANALYSIS},
	volume = {2020},
	issn = {2732-4494},
	doi = {10.1049/icp.2021.1252},
	pages = {121--126},
	number = {5},
	journaltitle = {{IET} Conference Proceedings},
	shortjournal = {{IET} Conf. Proc.},
	author = {Satyamsetti, V. K. and Michaelides, A. and Hadjiantonis, A.},
	date = {2021-06-02},
	langid = {english},

}

@inproceedings{kang_20_research_2025,
	location = {Zhenjiang, China},
	title = {Research on the Suppression Function of Reactive Power Compensation Device Based on Harmonic Source Detection},
	rights = {https://doi.org/10.15223/policy-029},
	isbn = {978-1-6654-7784-0},
	doi = {10.1109/PSPE66589.2025.11296395},
	eventtitle = {2025 4th International Conference on Power System and Power Engineering ({PSPE})},
	pages = {188--192},
	booktitle = {2025 4th International Conference on Power System and Power Engineering ({PSPE})},
	publisher = {{IEEE}},
	author = {Kang, Zhilin and Luo, Yuan and An, Jianzhen and Feng, Yanwei},
	date = {2025-09-19},
	langid = {english},
}

@inproceedings{jiang_12_quantification_2021,
	location = {Huzhou, China},
	title = {Quantification Index of Voltage Stability Based on Physical Mechanism of Reactive Power Distribution},
	rights = {https://ieeexplore.ieee.org/Xplorehelp/downloads/license-information/{IEEE}.html},
	isbn = {978-1-6654-3681-6},
	doi = {10.1109/SCEMS52239.2021.9646127},
	eventtitle = {2021 {IEEE} 4th Student Conference on Electric Machines and Systems ({SCEMS})},
	pages = {1--6},
	booktitle = {2021 {IEEE} 4th Student Conference on Electric Machines and Systems ({SCEMS})},
	publisher = {{IEEE}},
	author = {Jiang, Haoliang and Zhang, Hengxu},
	date = {2021-12-01},
	langid = {english},

}

@inproceedings{dondariya_15_voltage_2021,
	location = {Bhopal, India},
	title = {Voltage Stability Assessment and Improvement in Power Systems with Solar Photovoltaic Penetration},
	rights = {https://doi.org/10.15223/policy-029},
	isbn = {978-1-6654-0236-1},
	doi = {10.1109/ICEPES52894.2021.9699827},
	eventtitle = {2021 {IEEE} 2nd International Conference On Electrical Power and Energy Systems ({ICEPES})},
	pages = {1--4},
	booktitle = {2021 {IEEE} 2nd International Conference On Electrical Power and Energy Systems ({ICEPES})},
	publisher = {{IEEE}},
	author = {Dondariya, Chandrakant and Sakravdia, D.K.},
	date = {2021-12-10},
	langid = {english},

}

@article{ziegler_16_voltage_nodate,
	title = {Voltage Stability Enhancement by Reactive Power Changes based on Voltage Stability Index {PTSI}},
	author = {Ziegler, Mr Christian and Wolter, Martin},
	langid = {english},
}

@inproceedings{munyao_17_voltage_2021,
	location = {Nairobi, Kenya},
	title = {Voltage Stability Index Based on Multi-bus Reactive Power Loading},
	rights = {https://ieeexplore.ieee.org/Xplorehelp/downloads/license-information/{IEEE}.html},
	isbn = {978-1-6654-0311-5},
	doi = {10.1109/PowerAfrica52236.2021.9543435},
	eventtitle = {2021 {IEEE} {PES}/{IAS} {PowerAfrica}},
	pages = {1--5},
	booktitle = {2021 {IEEE} {PES}/{IAS} {PowerAfrica}},
	publisher = {{IEEE}},
	author = {Munyao, Peter M. and Agee, John T. and Tiako, Remy},
	date = {2021-08-23},
	langid = {english},

}

@article{yu_09_research_2022,
	title = {Research on Reactive Power Optimization Strategy under the Intelligent Improvement Model of the Distribution Network},
	volume = {2022},
	rights = {https://creativecommons.org/licenses/by/4.0/},
	issn = {1687-5699, 1687-5680},
	doi = {10.1155/2022/9310507},
	pages = {1--11},
	journaltitle = {Advances in Multimedia},
	shortjournal = {Advances in Multimedia},
	author = {Yu, Menglin},
	editor = {Li, Qiangyi},
	date = {2022-10-04},
	langid = {english},

}

@inproceedings{han_10_reactive_2022,
	location = {Shanghai, China},
	title = {Reactive power optimization control strategy for large-scale new energy access grid based on optimized particle swarm algorithm},
	rights = {https://doi.org/10.15223/policy-029},
	isbn = {978-1-6654-9899-9},
	eventtitle = {2022 4th International Conference on Electrical Engineering and Control Technologies ({CEECT})},
	pages = {1216--1221},
	booktitle = {2022 4th International Conference on Electrical Engineering and Control Technologies ({CEECT})},
	publisher = {{IEEE}},
	author = {Han, Xingning and Cai, Hui and Xu, Sixuan and Peng, Zhuyi and Zhao, Feifei and Qi, Wanchun},
	date = {2022-12},
	langid = {english},

}

@article{pudjianto_11_der_2020,
	title = {DER reactive services and distribution network losses},
	volume = {2020},
	issn = {2515-0855},
	doi = {10.1049/oap-cired.2021.0111},
	pages = {541--544},
	number = {1},
	journaltitle = {{CIRED} - Open Access Proceedings Journal},
	author = {Pudjianto, Danny and Djapic, Predrag and Strbac, Goran and Van Schalkwyk, Evert T. and Stojkovska, Biljana},
	date = {2020-01-01},
	langid = {english},

}

@inproceedings{xiaojiao_12_distribution_2022,
	location = {Xiamen, China},
	title = {Distribution network reactive power optimization considering multi flexible controlled resources},
	rights = {https://doi.org/10.15223/policy-029},
	isbn = {978-1-6654-6533-5},
	doi = {10.1109/CAC57257.2022.10055509},
	eventtitle = {2022 China Automation Congress ({CAC})},
	pages = {5587--5591},
	booktitle = {2022 China Automation Congress ({CAC})},
	publisher = {{IEEE}},
	author = {Xiaojiao, Liang and Juncheng, Si and Yuanyuan, Wang and Yanbin, Cai and Shuangle, Zhang and Yongchang, Guan},
	date = {2022-11-25},
	langid = {english},

}

@inproceedings{guo_14_reactive_2021,
	location = {Victoria, {BC}, Canada},
	title = {Reactive Power Optimization for Voltage Stability in Energy Internet Based on Graph Convolutional Networks and Deep Q-learning},
	rights = {https://ieeexplore.ieee.org/Xplorehelp/downloads/license-information/{IEEE}.html},
	isbn = {978-1-7281-6207-2},
	doi = {10.1109/ICPS49255.2021.9468118},
	eventtitle = {2021 4th {IEEE} International Conference on Industrial Cyber-Physical Systems ({ICPS})},
	pages = {511--516},
	booktitle = {2021 4th {IEEE} International Conference on Industrial Cyber-Physical Systems ({ICPS})},
	publisher = {{IEEE}},
	author = {Guo, Sheng and Cao, Junwei},
	date = {2021-05-10},
	langid = {english},

}

@inproceedings{song_01_comprehensive_2023,
	location = {Chengdu, China},
	title = {Comprehensive Configuration Method for Static and Dynamic Reactive Power Compensation of Multi-Voltage Level Urban Power Grid},
	rights = {https://doi.org/10.15223/policy-029},
	isbn = {979-8-3503-4506-3},
	doi = {10.1109/ICPES59999.2023.10400102},
	eventtitle = {2023 13th International Conference on Power and Energy Systems ({ICPES})},
	pages = {57--63},
	booktitle = {2023 13th International Conference on Power and Energy Systems ({ICPES})},
	publisher = {{IEEE}},
	author = {Song, Yunting and Gu, Jiming and Yuan, Zhenghai and Liu, Xinyu and Zhu, Shaoxuan and Yang, Gaofeng and Chen, Tao and Chen, Jialin},
	date = {2023-12-08},
	langid = {english},

}

@inproceedings{du_08_discussion_2010,
	location = {Chengdu, China},
	title = {Discussion and Researching on Dynamic Reactive Power Planning},
	isbn = {978-1-4244-4812-8},
	doi = {10.1109/APPEEC.2010.5448774},
	eventtitle = {2010 Asia-Pacific Power and Energy Engineering Conference},
	pages = {1--4},
	booktitle = {2010 Asia-Pacific Power and Energy Engineering Conference},
	publisher = {{IEEE}},
	author = {Du, Liang and Cheng, Fei},
	date = {2010},
	langid = {english},

}

@inproceedings{diaz_07_reactive_2009,
	location = {Cuernavaca, Morelos, Mexico},
	title = {Reactive Shunt Compensation Planning by Optimal Power Flows and Linear Sensitivities},
	isbn = {978-0-7695-3799-3},
	doi = {10.1109/CERMA.2009.73},
	eventtitle = {2009 Electronics, Robotics and Automotive Mechanics Conference},
	pages = {326--331},
	booktitle = {2009 Electronics, Robotics and Automotive Mechanics Conference ({CERMA})},
	publisher = {{IEEE}},
	author = {Diaz, Uriel A. Rangel and Hernandez, José H. Tovar},
	date = {2009-09},
	langid = {english},

}
%}
\end{document}